\makeatletter
\newcommand{\dontusepackage}[2][]{%
  \@namedef{ver@#2.sty}{9999/12/31}%
  \@namedef{opt@#2.sty}{#1}}
\makeatother
\dontusepackage{subfigure}

\documentclass[]{article}

\usepackage{lmodern}
\usepackage{amssymb,amsmath}
\usepackage{ifxetex,ifluatex}
\usepackage[usenames,dvipsnames]{color}
\usepackage{fixltx2e} % provides \textsubscript
\ifnum 0\ifxetex 1\fi\ifluatex 1\fi=0 % if pdftex
  \usepackage[T1]{fontenc}
  \usepackage[utf8]{inputenc}
\else % if luatex or xelatex
  \ifxetex
    \usepackage{mathspec}
    \usepackage{xltxtra,xunicode}
  \else
    \usepackage{fontspec}
  \fi
  \defaultfontfeatures{Mapping=tex-text,Scale=MatchLowercase}
  
\fi
\IfFileExists{upquote.sty}{\usepackage{upquote}}{}
\IfFileExists{microtype.sty}{%
\usepackage{microtype}
\UseMicrotypeSet[protrusion]{basicmath} % disable protrusion for tt fonts
}{}
\usepackage[margin=1.0in,bottom=1.5in]{geometry}
\usepackage[]{natbib}
\usepackage{listings}
\usepackage{graphicx}
\makeatletter
\def\maxwidth{\ifdim\Gin@nat@width>\linewidth\linewidth\else\Gin@nat@width\fi}
\def\maxheight{\ifdim\Gin@nat@height>\textheight\textheight\else\Gin@nat@height\fi}
\makeatother
\setkeys{Gin}{width=\maxwidth,height=\maxheight,keepaspectratio}
\usepackage{caption}
\usepackage{float}
\usepackage{subfig}
\ifxetex
  \usepackage[setpagesize=false, % page size defined by xetex
              unicode=false, % unicode breaks when used with xetex
              xetex]{hyperref}
\else
  \usepackage[unicode=true]{hyperref}
\fi
\hypersetup{breaklinks=true,
            bookmarks=true,
            pdfauthor={},
            pdftitle={De-risking geological carbon storage from high resolution time-lapse seismic to explainable leakage detection},
            colorlinks=true,
            citecolor=black,
            urlcolor=blue,
            linkcolor=black,
            pdfborder={0 0 0}}
\title{De-risking geological carbon storage from high resolution time-lapse
seismic to explainable leakage detection}
\author{Ziyi Yin, Huseyin Tuna Erdinc, Abhinav Prakash Gahlot, Mathias
Louboutin, Felix J. Herrmann\\Georgia Institute of Technology\\}
\date{}

\begin{document}
\maketitle
\begin{abstract}
Geological carbon storage represents one of the few truly scalable
technologies capable of reducing the CO\textsubscript{2} concentration
in the atmosphere. While this technology has the potential to scale, its
success hinges on our ability to mitigate its risks. An important aspect
of risk mitigation concerns assurances that the injected
CO\textsubscript{2} remains within the storage complex. Amongst the
different monitoring modalities, seismic imaging stands out with its
ability to attain high resolution and high fidelity images. However,
these superior features come, unfortunately, at prohibitive costs and
time-intensive efforts potentially rendering extensive seismic
monitoring undesirable. To overcome this shortcoming, we present a
methodology where time-lapse images are created by inverting
non-replicated time-lapse monitoring data jointly. By no longer
insisting on replication of the surveys to obtain high fidelity
time-lapse images and differences, extreme costs and time-consuming
labor are averted. To demonstrate our approach, hundreds of noisy
time-lapse seismic datasets are simulated that contain imprints of
regular CO\textsubscript{2} plumes and irregular plumes that leak. These
time-lapse datasets are subsequently inverted to produce time-lapse
difference images used to train a deep neural classifier. The testing
results show that the classifier is capable of detecting
CO\textsubscript{2} leakage automatically on unseen data and with a
reasonable accuracy.
\end{abstract}

\section{Introduction}\label{introduction}

For various reasons, seismic monitoring of geological carbon storage
(GCS) comes with its own set of unique challenges. Amongst these
challenges, the need for low-cost highly repeatable, high resolution,
and high fidelity images ranks chiefly. While densely sampled and
replicated time-lapse surveys---which rely on permanent reservoir
monitoring systems---may be able to provide images conducive to
interpretation and reservoir management, these approaches are often too
costly and require too much handholding to be of practical use for GCS.

To overcome these challenges, we replace the current paradigm of costly
replicated acquisition, cumbersome time-lapse processing, and
interpretation, by a joint inversion framework mapping time-lapse data
to high fidelity and high resolution images from sparse non-replicated
time-lapse surveys. We demonstrate that we arrive at an imaging
framework that is suitable for automatic detection of pressure-induced
CO\textsubscript{2} leakage. Rather than relying on meticulous 4D
workflows where baseline and monitoring surveys are processed separately
to yield accurate and artifact-free time-lapse differences, our approach
exposes information that is shared amongst the different vintages by
formulating the imaging problem in terms of an unknown fictitious common
component, and innovations of the baseline and monitor survey(s) with
respect to this common component. Because the common component is
informed by all time-lapse surveys, its image quality improves when the
surveys bring complementary information, which is the case when the
surveys are not replicated. In turn, the enhanced common component
results in improved images for the baseline, monitor, and their
time-lapse difference. Joint inversion also leads to robustness with
respect to noise, calibration errors, and time-lapse changes in the
background velocity model.

To showcase the achievable imaging gains and how these can be used in a
GCS setting where CO\textsubscript{2} leakage is of major consideration,
we create hundreds of time-lapse imaging experiments involving
CO\textsubscript{2} plumes whose behavior is determined by the two-phase
flow equations. To mimic irregular flow due to pressure-induced opening
of fractures, we increase the permeability in the seal at random
locations and pressure thresholds. The resulting flow simulations are
used to generate time-lapse datasets that serve as input to our joint
imaging scheme. The produced time-lapse difference images are
subsequently used to train and test a neural network that as an
explainable classifier determines whether the CO\textsubscript{2} plume
behaves regularly or shows signs of leakage.

Our contributions are organized as follows. First, we discuss the
time-lapse seismic imaging problem and its practical difficulties. Next,
we introduce the joint recovery model that takes explicit advantage of
information shared by multiple surveys. By means of a carefully designed
synthetic case study involving saline aquifers made of Blunt sandstone
in the Southern North Sea, we demonstrate the uplift of the joint
recovery model and how its images can be used to train a deep neural
network classifier to detect erroneous growth of the CO\textsubscript{2}
plume automatically. Aside from determining whether the
CO\textsubscript{2} plume behaves regularly or not, our network also
provides class activation mappings that visualize areas in the image on
which the network is basing its classification.

\section{Seismic monitoring with time-lapse
imaging}\label{seismic-monitoring-with-time-lapse-imaging}

To keep track of CO\textsubscript{2} plume development during geological
carbon storage (GCS) projects, multiple time-lapse surveys are
collected. Baseline surveys are acquired before the supercritical
CO\textsubscript{2} is injected into the reservoir. These baseline
surveys, denoted by the index $j=1$, are followed by one or more monitor
surveys, collected at later times and indexed by $j=2,\cdots,n_v$ with
$n_v$ the total number of surveys.

Seismic monitoring of GCS brings its own unique set of challenges that
stem from the fact that its main concern is (early) detection of
possible leakage of CO\textsubscript{2} from the storage complex. To be
successful with this task, monitoring GCS calls for a time-lapse imaging
modality that is capable of

\begin{itemize}
\itemsep1pt\parskip0pt\parsep0pt
\item
  detecting weak time-lapse signals associated with small rock-physics
  changes induced by CO\textsubscript{2} leakage
\item
  attaining high lateral resolution from active-source surface seismic
  data to detect vertically moving leakage
\item
  handling an increasing number of seismic surveys collected over long
  periods of time ($\sim50-100$ years)
\item
  reducing costs drastically by no longer insisting on replication of
  time-lapse surveys to attain high degrees of repeatability
\item
  lowering the cumulative environmental imprint of active source
  acquisition
\end{itemize}

\subsection{Monitoring with the joint recovery
model}\label{monitoring-with-the-joint-recovery-model}

To meet these challenges, we choose a linear imaging framework where
observed linearized data for each vintage is related to perturbations in
the acoustic impedance via
\begin{equation}
\mathbf{b}_j=\mathbf{A}_j\mathbf{x}_j+\mathbf{e}_j\quad \text{for}\quad j=1,2,\cdots,n_v.
\label{eq-lin-model}
\end{equation}
 In this expression, the matrix $\mathbf{A}_j$ stands for the linearized
Born scattering operator for the $j\,\mathrm{th}$ vintage. Observed
linearized data, collected for all shots in the vector $\mathbf{b}_j$,
is generated by applying the $\mathbf{A}_j$'s to the (unknown) impedance
perturbations denoted by $\mathbf{x}_j$ for $j=1,2,\cdots, n_v$. The
task of time-lapse imaging is to create high resolution, high fidelity
and true amplitude estimates for the time-lapse images,
$\left\{\widehat{\mathbf{x}}_j\right\}_{j=1}^{n_v}$, from non-replicated
sparsely sampled noisy time-lapse data.

We argue that our choice for linearized imaging is justified for four
reasons. First, CO\textsubscript{2}-injection sites undergo extensive
baseline studies, which means that accurate information on the
background velocity model is generally available. Second, changes in the
acoustic parameters induced by CO\textsubscript{2} injection are
typically small, so it suffices to work with one and the same background
model for the baseline and monitor surveys. Third, when the background
model is sufficiently close to the true model, linearized inversion,
which corresponds to a single Gauss-Newton iteration of full-waveform
inversion, converges quadratically. Fourth, because the forward model is
linear, it is conducive to the use of the joint recovery model where
inversions are carried out with respect to the common component, which
is shared between all vintages, and innovations with respect to the
common component. Because the common component represents an average, we
expect this joint imaging method to be relatively robust with respect to
kinematic changes induced by time-lapse effects or by lack of
calibration of the acquisition \citep{oghenekohwo2017hrt}.

By parameterizing time-lapse images,
$\left\{\mathbf{x}_j\right\}_{j=1}^{n_v}$, in terms of the common
component, $\mathbf{z}_0$, and innovations with respect to the common
component, $\left\{\mathbf{z}_j\right\}_{j=1}^{n_v}$, we arrive at the
joint recovery model where representations for the images are given by
\begin{equation}
\mathbf{x}_j = \frac{1}{\gamma}\mathbf{z}_0 + \mathbf{z}_j \quad \text{for} \quad j=1,2,\cdots,n_v.
\label{eq-components}
\end{equation}
 Here, the parameter, $\gamma$, controls the balance between the common
component, $\mathbf{z}_0$, and innovation components,
$\left\{\mathbf{z}_j\right\}_{j=1}^{n_v}$ \citep{li2015weighted}.
Compared to traditional time-lapse approaches, where data are imaged
separately or where time-lapse surveys are subtracted, inversions for
time-lapse images based on the above parameterization are carried out
jointly and involve inverting the following matrix:
\begin{equation}
\begin{aligned}
\mathbf{A} = \begin{bmatrix} 
\frac{1}{\gamma} \mathbf{A}_1 & \mathbf{A}_1 &  & \\ 
\vdots & & \ddots &   \\
\frac{1}{\gamma} \mathbf{A}_{n_v} & &  & \mathbf{A}_{n_v}
\end{bmatrix}.
\end{aligned}
\label{eq-jrm}
\end{equation}
 While traditional time-lapse imaging approaches strive towards maximal
replication between the surveys to suppress acquisition related
artifacts, imaging with the joint recovery model---which entails
inverting the underdetermined system in Equation~\ref{eq-jrm} using
structure-promotion techniques (e.g.~via $\ell_1$-norm
minimization)---improves the image quality of the vintages themselves in
situations where the surveys are not replicated. This occurs in cases
where $\mathbf{A}_i\neq\mathbf{A}_j$ for $\forall i\neq j$, or in
situations where there is significant noise. This remarkable result was
shown to hold for sparsity-promoting denoising of time-lapse field data
\citep{wei2018improve, tian2018joint}, for various wavefield
reconstructions of randomized simultaneous-source dynamic (towed-array)
and static (OBC/OBN) marine acquisitions
\citep{oghenekohwo2017hrt, kotsi2020time, zhou2021non}, and for
wave-based inversion, including least-squares reverse-time migration and
full-waveform inversion
\citep{oghenekohwo2017THetl, oghenekohwo2017EAGEitl}. The observed
quality gains in these applications can be explained by improvements in
the common component resulting from complementary information residing
in non-replicated time-lapse surveys. This enhanced recovery of the
common component in turn improves the recovery of the innovations and
therefore the vintages themselves. The time-lapse differences themselves
also improve, or at the very least, remain relatively unaffected when
the surveys are not replicated. Relaxing replication of surveys
obviously leads to reduction in cost and environmental impact. Below, we
show how GCS monitoring also benefits from this approach.

\subsection{Monitoring with curvelet-domain structure
promotion}\label{monitoring-with-curvelet-domain-structure-promotion}

To obtain high resolution and high fidelity time-lapse images, we invert
the system in Equation~\ref{eq-jrm}
\citep{yang2020tdsp, witte2018cls, yin2021SEGcts} with
\begin{equation}
\begin{split}
\underset{\mathbf{z}}{\operatorname{minimize}} \quad \lambda \|\mathbf{C}\mathbf{z}\|_1+\frac{1}{2}\|\mathbf{C}\mathbf{z}\|_2^2 \\
\text{subject to}\quad \|\mathbf{b}- \mathbf{A}\mathbf{z}\|_2^2 \leq \sigma,
\end{split}
\label{eq-elastic}
\end{equation}
 where $\mathbf{C}$ is the forward curvelet transform, $\lambda$ the
threshold parameter, and $\sigma$ the magnitude of the noise. At
iteration $k$ and for $\sigma=0$, solving Equation~\ref{eq-elastic}
corresponds to computing the following iterations:
\begin{equation}
\begin{aligned}
\begin{array}{lcl} 
  \mathbf{u}_{k+1} & = & \mathbf{u}_k-t_k \mathbf{A}_k^\top(\mathbf{A}_k\mathbf{z}_{k}-\mathbf{b}_k)\\
 \mathbf{z}_{k+1} & = & \mathbf{C}^\top S_{\lambda}(\mathbf{C}\mathbf{u}_{k+1}),
\end{array}
\end{aligned}
\label{LBk}
\end{equation}
 where $\mathbf{A}_k$, with a slight abuse of notation, represents the
matrix in Equation~\ref{eq-jrm} for a subset of shots randomly selected
from sources in each survey. The vector $\mathbf{b}_k$ contains the
extracted shot records from $\mathbf{b}$ and the symbol $^\top$ refers
to the adjoint. Sparsity is promoted via curvelet-domain soft
thresholding,
$S_{\lambda}(\cdot)=\max(|\cdot|-\lambda,0)\operatorname{sign}(\cdot)$,
where, $\lambda$, is the threshold. The vectors $\mathbf{u}_k$ and
$\mathbf{z}_k$ contain the baseline and innovation components.

\section{Numerical case study: Blunt sandstone in the Southern North
Sea}\label{numerical-case-study-blunt-sandstone-in-the-southern-north-sea}

\begin{figure}
\centering
\includegraphics[width=0.980\hsize]{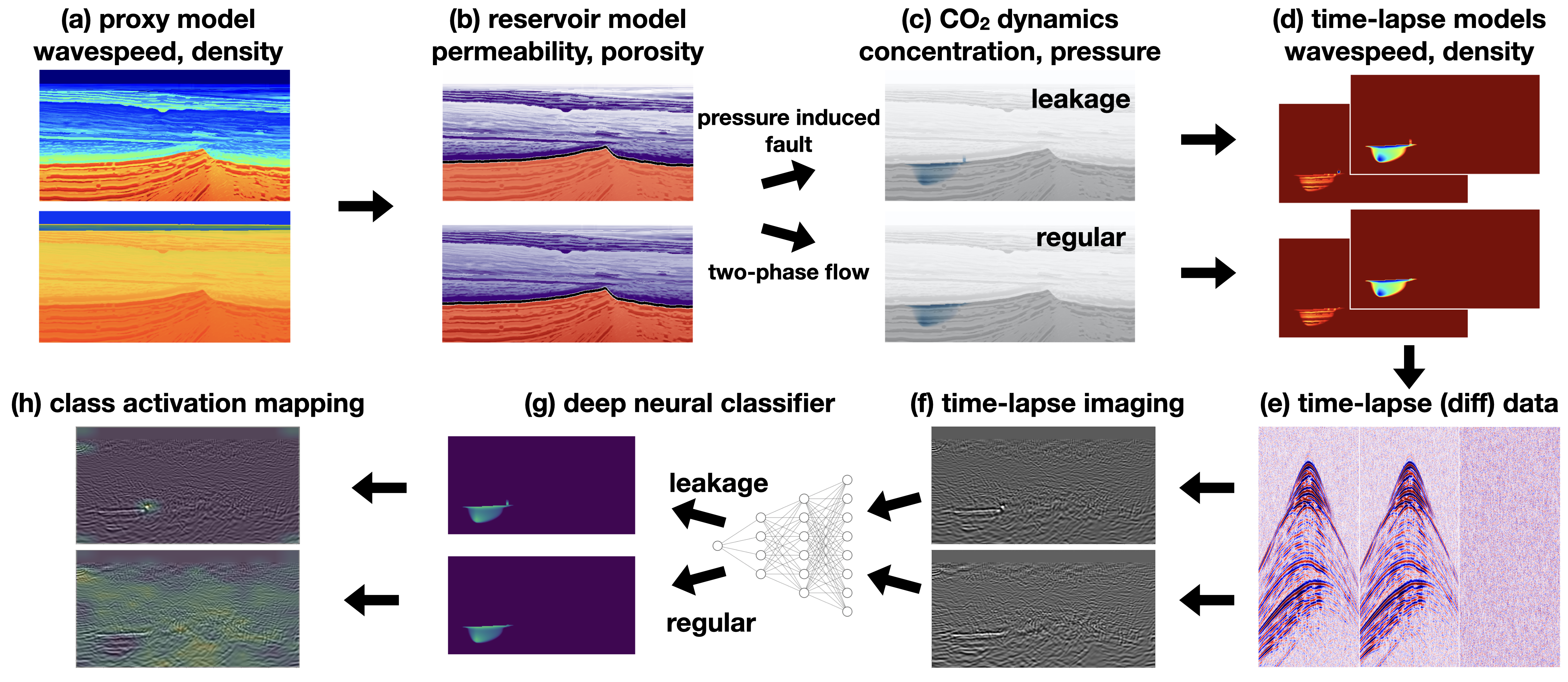}
\caption{Simulation-based monitoring design framework. Starting with a
proxy model for the wavespeed and density (a), the workflow proceeds by
converting these seismic properties into permeability and porosity (b).
These fluid flow properties are used to simulated CO\textsubscript{2}
plumes that behave regularly or exhibit leakage outside the storage
complex (c). Induced changes by the CO\textsubscript{2} plume for the
wavespeed and density are depicted in (d) and serve as input to
simulations of time-lapse seismic data (SNR $8.0\,\mathrm{dB}$) and
shot-domain time-lapse differences (SNR $-31.4\,\mathrm{dB}$). Imaging
results for regular and irregular plume developments are plotted in (f)
and serve as input to the deep neural classifier (g), which determines
whether the flow behaves regularly or leaks. Activation mappings in (h)
show regions on which the network is basing its classification. As
expected, the activation mapping is diffusive in case of regular
CO\textsubscript{2} plume development and focused on the leakage
location when CO\textsubscript{2} plume behaves
irregularly.}\label{fig:workflow}
\end{figure}

Before discussing the impact of high resolution and high fidelity
time-lapse imaging with the joint recovery model on the down-stream task
of automatic leakage detection with a neural network classifier, we
first detail the setup of our numerical experiments using techniques
from simulation-based acquisition design as described by
\citet{yin2021SEGcts}. In order to generate realistic time-lapse data
and training sets for the automatic leakage classifier, we follow the
workflow summarized in Figure~\ref{fig:workflow}. In this approach, use
is made of proxy models for seismic properties derived from real 3D
imaged seismic and well data \citep{jones2012building}. With rock
physics, these seismic models are converted to fluid-flow models that
serve as input to two-phase flow simulations. The resulting datasets,
which include pressure-induced leakage, will be used to create
time-lapse data used to train our classifier. For more detail, refer to
the caption of Figure~\ref{fig:workflow}.

\subsection{Proxy seismic and fluid-flow
models}\label{proxy-seismic-and-fluid-flow-models}

\begin{figure}
\centering
\subfloat[\label{fig:permeability}]{\includegraphics[width=0.490\hsize]{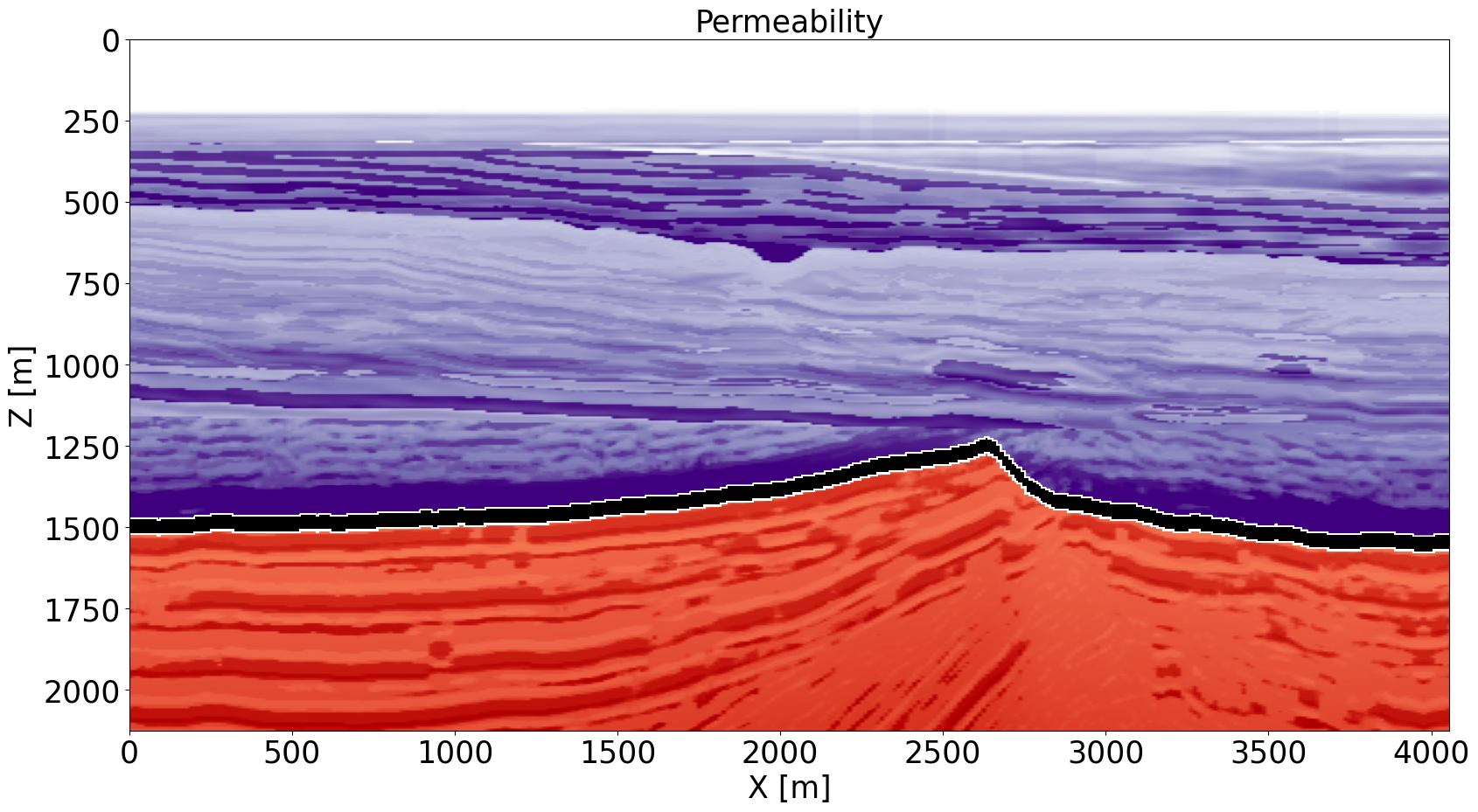}}
\subfloat[\label{fig:porosity}]{\includegraphics[width=0.490\hsize]{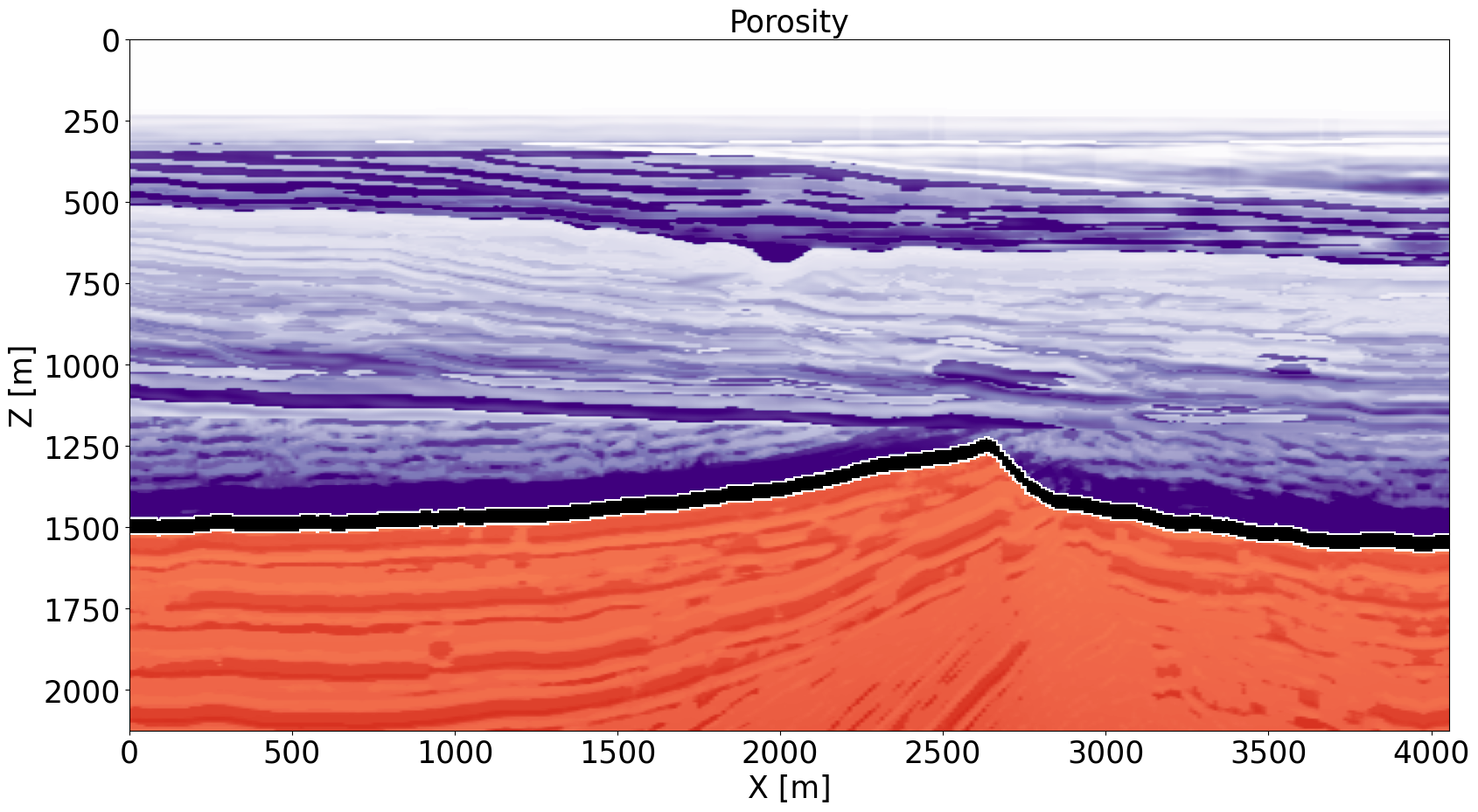}}
\caption{Permeability and porosity derived from a 2D slice of Compass
model.}\label{fig:perm}
\end{figure}

Amongst the various CO\textsubscript{2} injection projects, GCS in
offshore saline aquifers has been most successful in reaching scale and
in meeting injection targets. For that reason, we consider a proxy model
constructed representative for CO\textsubscript{2} injection in the
South of the North Sea involving a saline aquifer made of the highly
permeable Blunt sandstone. This area, which is actively being considered
for GCS \citep{kolster2018impact}, consists of the following three
geologic sections (see Figure~\ref{fig:perm} for the permeability and
porosity distribution):

\begin{enumerate}
\def\labelenumi{(\roman{enumi})}
\item
  the highly porous (average $33\%$) and permeable ($>170\,\mathrm{mD}$)
  Blunt sandstone reservoir of about $300-500\,\mathrm{m}$ thick. This
  section, denoted by red colors in Figure~\ref{fig:perm}, corresponds
  to the saline aquifer and serves as the reservoir for
  CO\textsubscript{2} injection;
\item
  the primary seal (permeability $10^{-4}-10^{-2}\,\mathrm{mD}$) made of
  the Rot Halite Member, which is $50\,\mathrm{m}$ thick and continuous
  (black layer in Figure~\ref{fig:perm});
\item
  the secondary seal made of the Haisborough group, which is
  $>300\,\mathrm{m}$ thick and consists of low-permeable (permeability
  $15-18\,\mathrm{mD}$) mudstone (purple section in
  Figure~\ref{fig:perm}).
\end{enumerate}

To arrive at the fluid-flow models, we consider 2D subsets of the 3D
Compass model \citep{jones2012building} and convert these seismic models
to fluid-flow properties (see Figure~\ref{fig:workflow} (b)) by assuming
a linear relationship between compressional wavespeed and permeability
in each stratigraphic section. For further details on the conversion of
compressional wavespeed and density to permeability and porosity, we
refer to empirical relationships reported in
\citet{klimentos1991effects}. During conversion, an increase of
$1\mathrm{km/s}$ in compressional wavespeed is assumed to correspond to
an increase of $1.63\,\mathrm{mD}$ in permeability. From this, porosity
is calculated with the Kozeny-Carman equation
\citep{costa2006permeability}
$K = \mathbf{\phi}^3 \left(\frac{1.527}{0.0314*(1-\mathbf{\phi})}\right)^2$,
where $K$ and $\phi$ denote permeability ($\mathrm{mD}$) and porosity
(\%) with constants taken from the Strategic UK CCS Storage Appraisal
Project report.

\subsection{Fluid-flow simulations}\label{fluid-flow-simulations}

To model CO\textsubscript{2} plumes that behave regularly and
irregularly, the latter due to leakage, we solve the two-phase flow
equations numerically\footnote{We used the open-source software
  \href{https://github.com/lidongzh/FwiFlow.jl}{FwiFlow.jl}
  \citep{kailaix_2021_5528428, li2020coupled} to solve the two-phase
  flow equations for both the pressure and concentration.} for both
pressure and concentration \citep{kailaix_2021_5528428, li2020coupled}.
To mimic possible pressure-induced CO\textsubscript{2} leakage, we
increase the permeability at random distances away from the injection
well within the seal from $10^{-4}\,\mathrm{mD}$ to $500\,\mathrm{mD}$
when the pressure exceeds $\sim 15\,\mathrm{MPa}$. At that depth, the
pressure is below the fracture gradient \citep{ringrose2020store}. Since
pressure-induced fractures come in different sizes, we also randomly
vary the width of the pressure-induced fracture openings from
$12.5\,\mathrm{m}$ to $62.5\,\mathrm{m}$. Examples of fluid-flow
simulations without and with leakage are shown in
Figure~\ref{fig:workflow} (c).

\subsection{Rock-physics conversion}\label{rock-physics-conversion}

To monitor temporal variations in the plume's CO\textsubscript{2}
concentration seismically, we use the patchy saturation model
\citep{avseth2010quantitative} to convert the CO\textsubscript{2}
concentration to decrease in compressional wavespeed and density. These
changes are shown in Figure~\ref{fig:workflow} from (d). The fact that
these induced changes in the time-lapse differences in seismic
properties are relatively small in spatial extent
($\sim 800\,\mathrm{m}$ for the plume and $< 62.5 \,\mathrm{m}$ for the
leakage) and amplitude ($18\%$ of the impedance perturbations with
respect to the background model) calls for a time-lapse imaging modality
with small normalized root-mean-square (NRMS) \citep{kragh2002seismic}
values $(\sim 2.5\%)$. This NMRS value is based on the relative
amplitude of impedance perturbations with respect to the background
model ( $14\%$).

\subsection{Time-lapse seismic
simulations}\label{time-lapse-seismic-simulations}

To train and validate automatic detection of CO\textsubscript{2} leakage
from the storage complex requires the creation of realistic time-lapse
datasets that contain the seismic imprint of regular as well as
irregular (leakage) plume development. To this end, baseline surveys are
simulated prior to CO\textsubscript{2} injection for different subsets
of the Compass model. Monitor surveys are simulated $200$ days after
leakage occurs to verify that potential leakage can be detected
automatically early on. For regular plume development, we shoot monitor
surveys for each subset at random times after CO\textsubscript{2}
injection. In order to strike a balance between acquisition productivity
and time-lapse image quality, use is made of dense permanent acoustic
monitoring at the seafloor with $25\,\mathrm{m}$ receiver spacing.
Time-lapse acquisition costs are reduced by non-replicated coarse
shooting with the source towed at $10\,\mathrm{m}$ depth below the ocean
surface. Subsampling artifacts are reduced by using a randomized
technique from compressive sensing where $32$ sources are located at
non-replicated jittered \citep{herrmann2008GJInps} source positions,
yielding an average source sampling of $125\,\mathrm{m}$. Given this
acquisition geometry, linear data is generated\footnote{We used the
  open-source software
  \href{https://github.com/slimgroup/JUDI.jl}{JUDI.jl}
  \citep{witte2018alf, mathias_louboutin_2022_7086719} to model the wave
  propagation. This Julia package calls the highly optimized propagators
  of \href{https://www.devitoproject.org/}{Devito}
  \citep{louboutin2018dae, luporini2020architecture, fabio_luporini_2022_6958070}.}
with Equation~\ref{eq-lin-model} for a $25\,\mathrm{Hz}$ Ricker wavelet
and with the band-limited noise term set so that the data's
signal-to-noise ratio (SNR) is $8.0\,\mathrm{dB}$. This noise level
leads to an extremely poor SNR of $-31.4\,\mathrm{dB}$ for time-lapse
differences in the shot domain. See Figure~\ref{fig:workflow} (e).

\subsection{Imaging with joint recovery model versus reverse-time
migration}\label{imaging-with-joint-recovery-model-versus-reverse-time-migration}

\begin{figure}
\centering
\subfloat[\label{fig:RTM}]{\includegraphics[width=0.490\hsize]{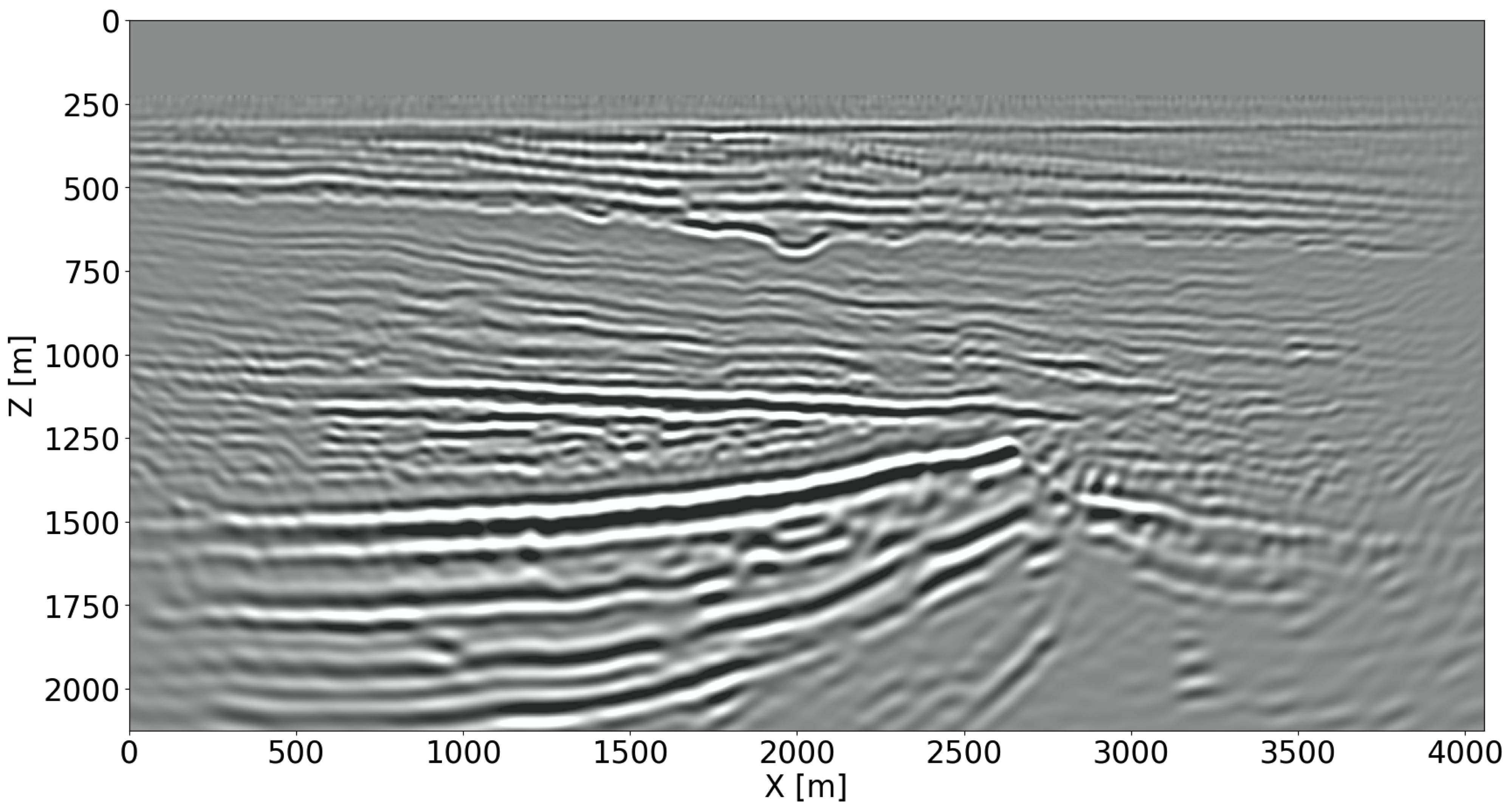}}
\subfloat[\label{fig:JRM}]{\includegraphics[width=0.490\hsize]{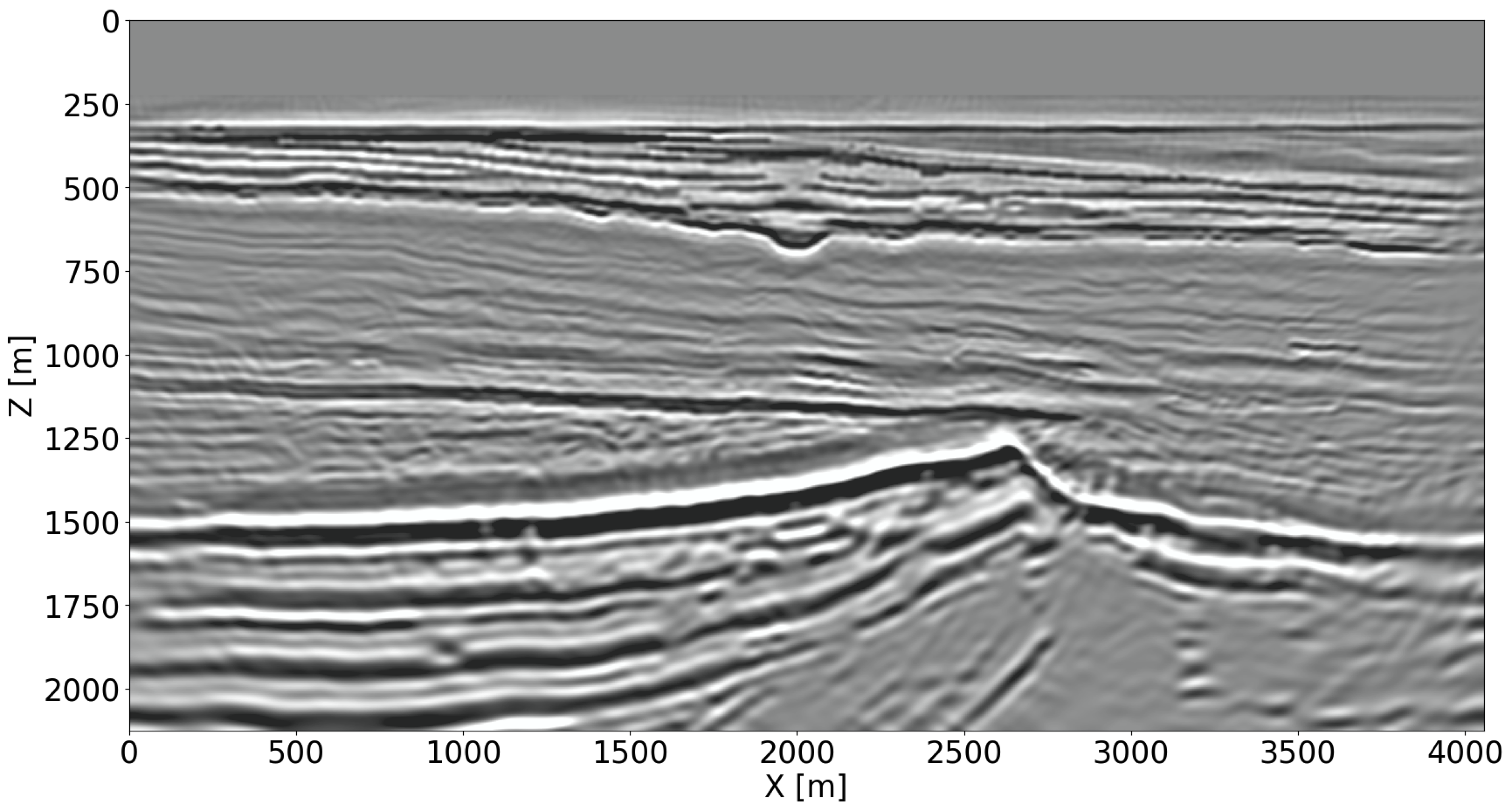}}
\\
\subfloat[\label{fig:diff_RTM}]{\includegraphics[width=0.490\hsize]{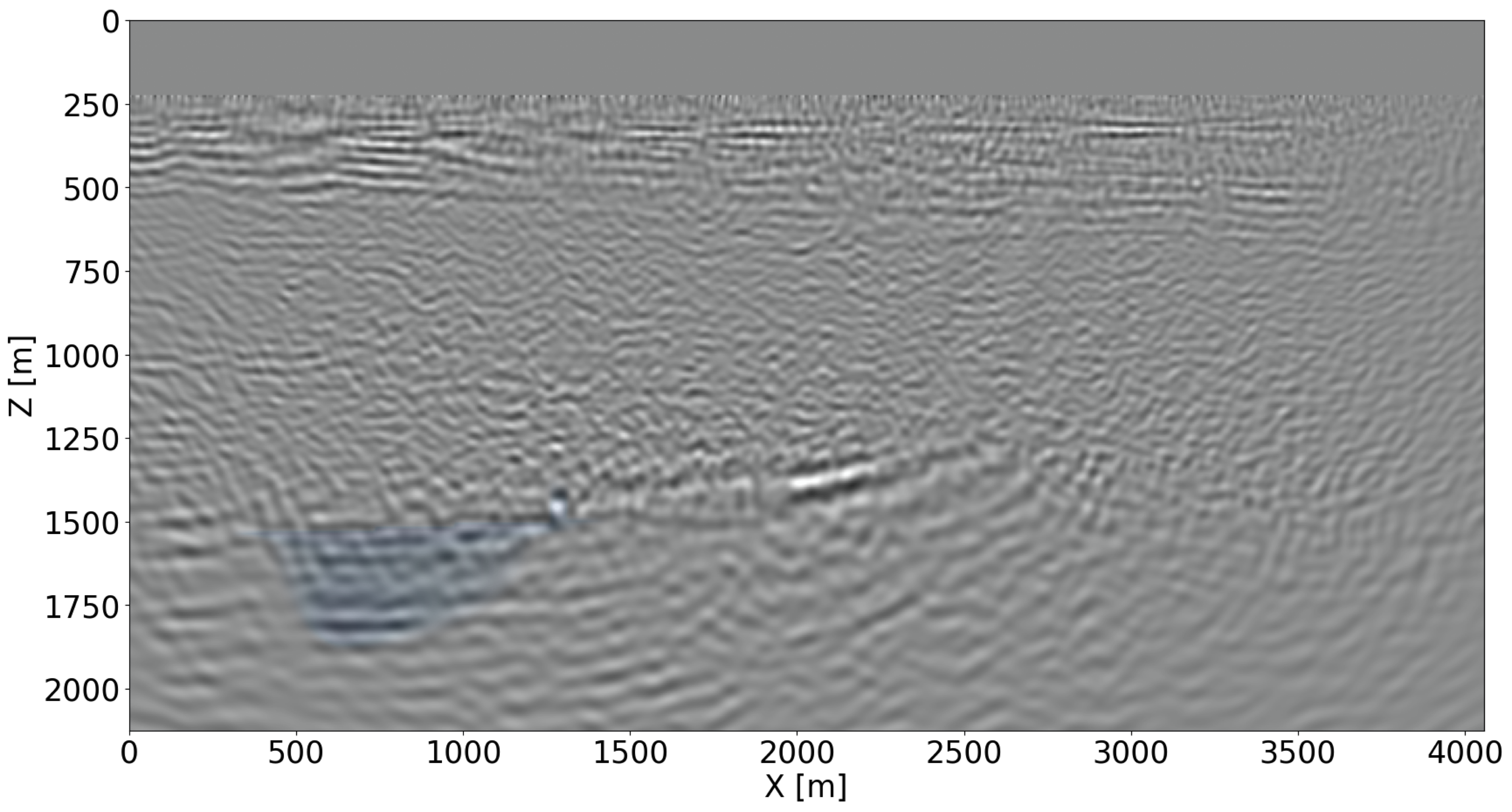}}
\subfloat[\label{fig:diff_JRM}]{\includegraphics[width=0.490\hsize]{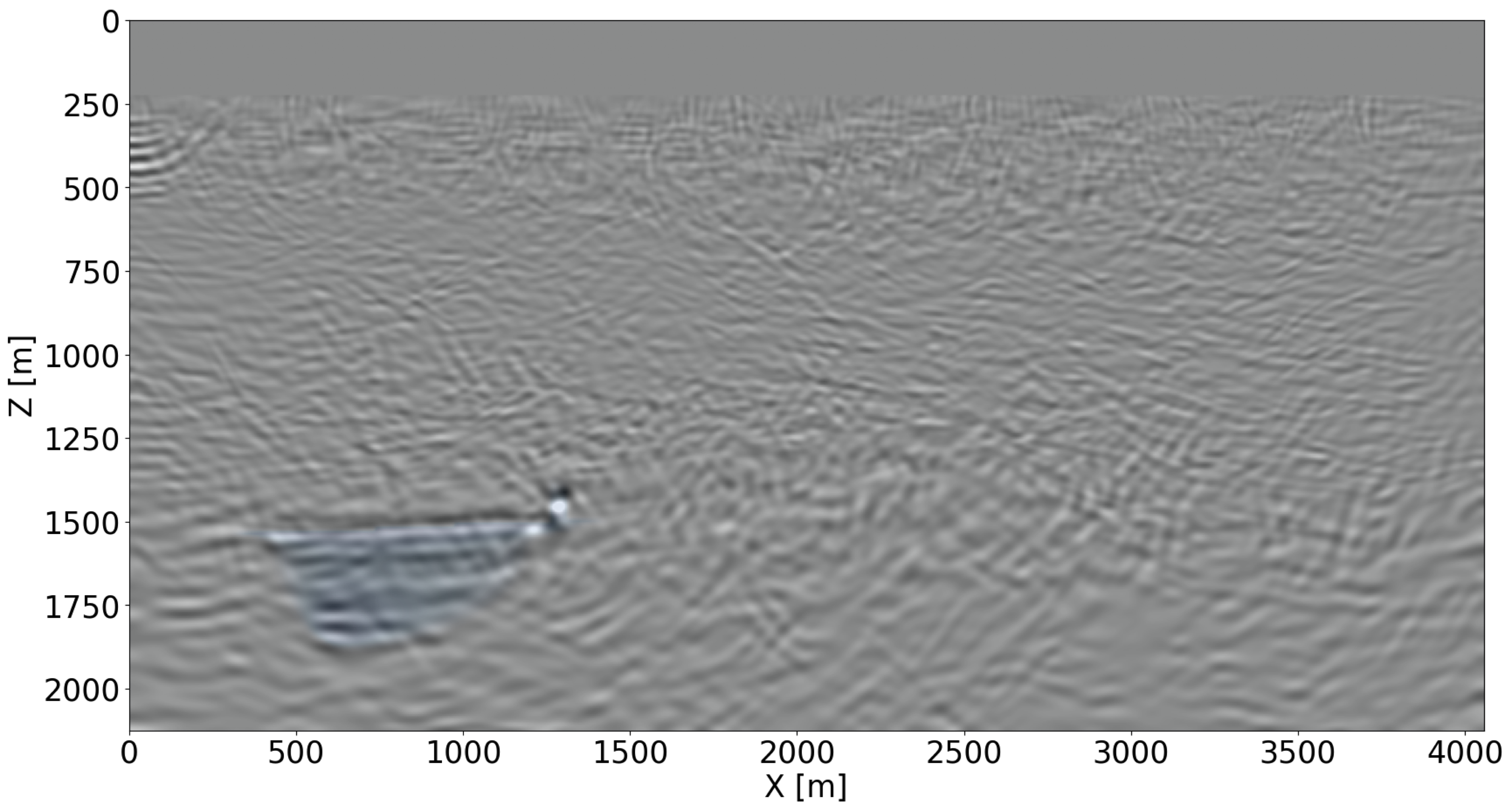}}
\\
\subfloat[\label{fig:diff_RTM_noleak}]{\includegraphics[width=0.490\hsize]{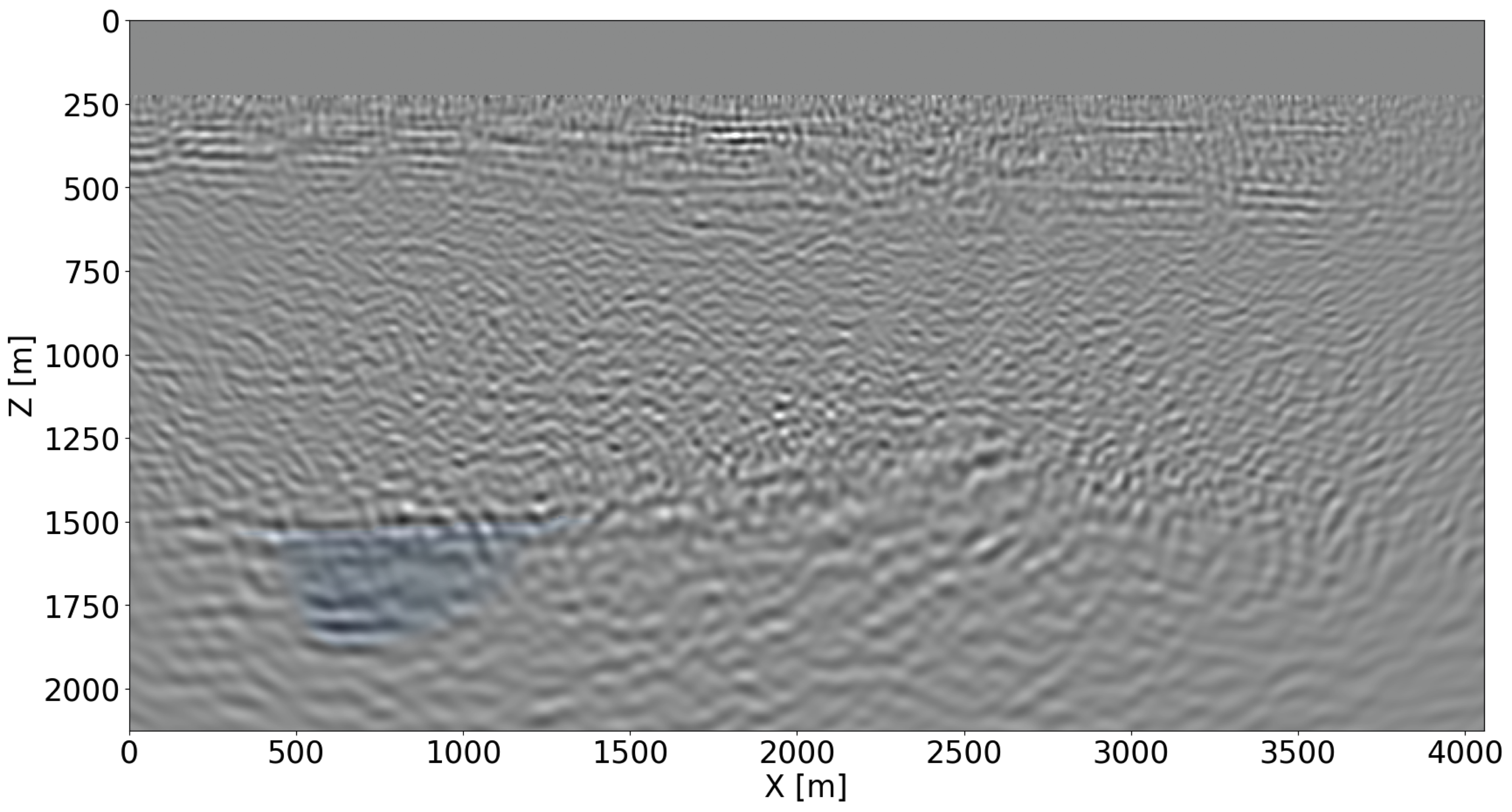}}
\subfloat[\label{fig:diff_JRM_noleak}]{\includegraphics[width=0.490\hsize]{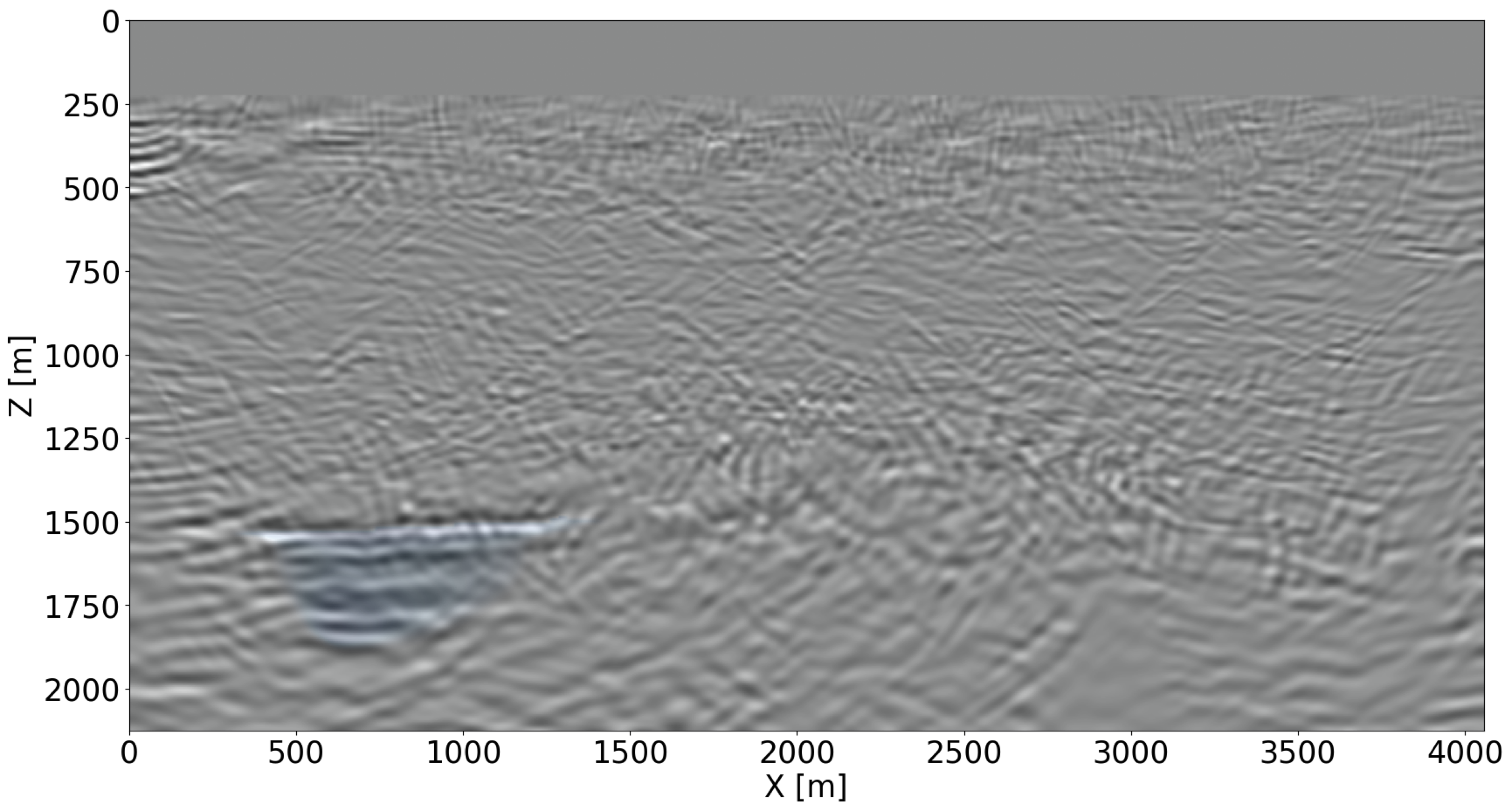}}
\caption{Reverse-time migration (RTM) versus inversion joint recovery
model (JRM). (a) RTM image of the baseline; (b) JRM image of the
baseline; (c) time-lapse difference and CO\textsubscript{2} plume for
independent RTM images with leakage; (d) time-lapse obtained by
inverting the time-lapse data jointly with leakage; (e) time-lapse
difference and CO\textsubscript{2} plume for independent RTM images
without leakage; (f) time-lapse obtained by inverting the time-lapse
data jointly without leakage. Notice improvement in the time-lapse image
quality. This improvement in reflected in the NRMS values that decrease
from $8.48\,\%$ for RTM to $3.20\, \%$ for
JRM.}\label{fig-difference_RTM_vs_JRM}
\end{figure}

\begin{figure}
\centering
\includegraphics[width=0.980\hsize]{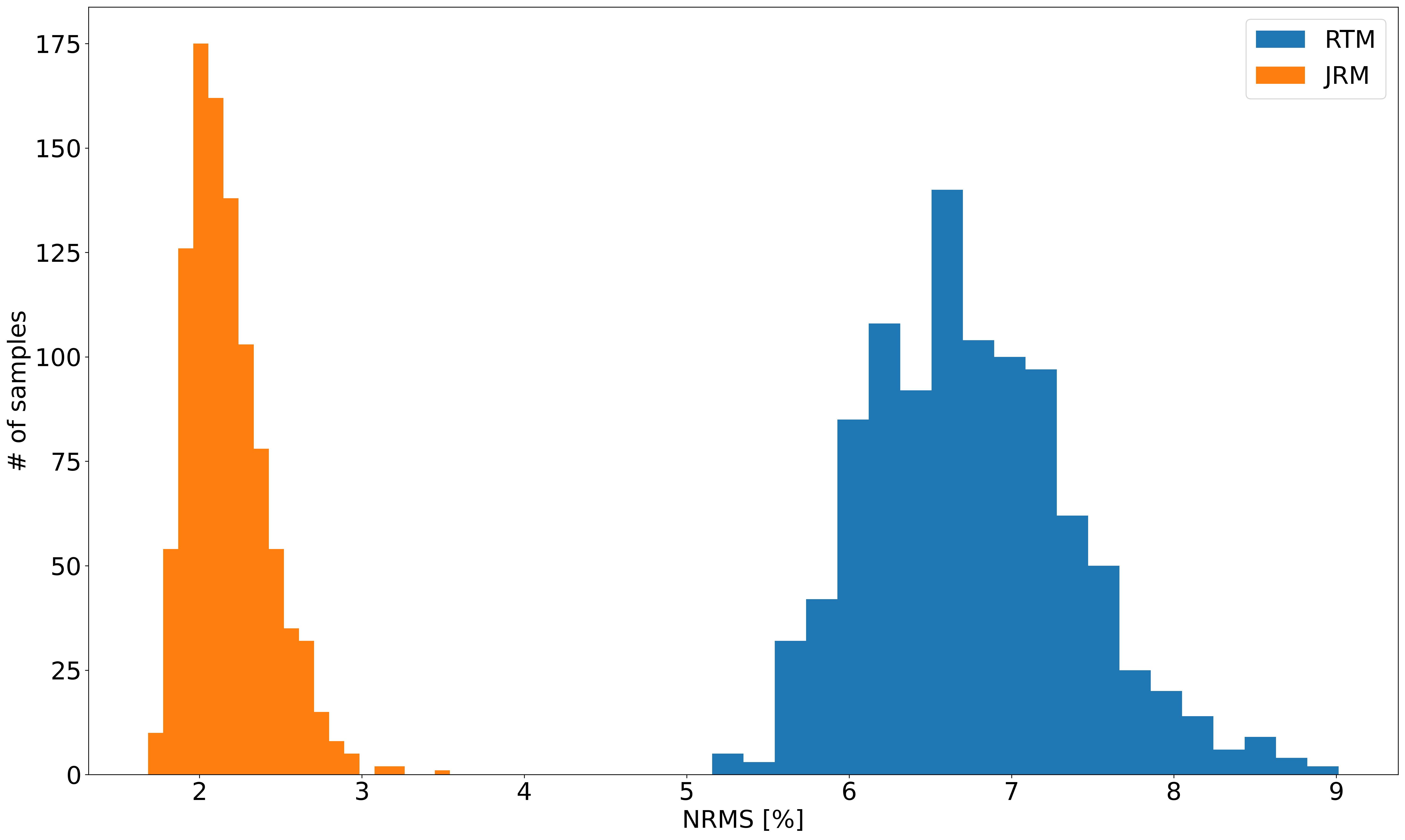}
\caption{NRMS values for $1000$ time-lapse
experiments.}\label{fig-difference_hist_RTM_vs_JRM}
\end{figure}

Given the simulated time-lapse datasets with and without leakage,
time-lapse difference images are created according to two different
imaging scenarios, namely via independent reverse-time migration (RTM),
conducted on the baseline and monitor surveys separately, and via
inversion of the joint recovery model (cf.~Equations~\ref{eq-jrm}
and~\ref{eq-elastic}). To limit the computational cost of the Bregman
iterations (Equation~\ref{LBk}), four shot records are selected per
iteration at random from each survey for imaging
\citep{yin2008bregman, witte2018cls, yang2020tdsp, yin2021SEGcts},
limiting the cost of the joint inversion to three RTMs. The recovered
baseline images are shown in Figures~\ref{fig:RTM} for RTM
and~\ref{fig:JRM} for JRM. For the leakage scenario, the time-lapse
differences are plotted in Figures~\ref{fig:diff_RTM}
and~\ref{fig:diff_JRM}, for RTM and JRM respectively. For the regular
plume, the time-lapse differences are plotted in
Figures~\ref{fig:diff_RTM_noleak} and~\ref{fig:diff_JRM_noleak}, for RTM
and JRM respectively. From these images, it is clear that joint
inversion leads to relatively artifact-free recovery of the vintages and
time-lapse differences. This observation is reflected in the NRMS
values, which improve considerably as shown by the histograms in
Figure~\ref{fig-difference_hist_RTM_vs_JRM}~for $1000$ imaging
experiments. Not only do the NRMS values shift towards the left, their
values are also more concentrated when inverting time-lapse data with
the joint recovery model. Both features are beneficial to automatic
leakage detection.

\section{\texorpdfstring{Deep neural network classifier for
CO\textsubscript{2} leakage
detection}{Deep neural network classifier for CO2 leakage detection}}\label{deep-neural-network-classifier-for-co2-leakage-detection}

The injection of supercritical CO\textsubscript{2} into the storage
complex perturbs the physical, chemical and thermal environment of the
reservoir \citep{newell2019overview}. Because CO\textsubscript{2}
injection increases the pressure, this process may trigger
CO\textsubscript{2} leakage across the seal when the pressure increase
induces opening of pre-existing faults or fractures zones
\citep{pruess2006co2, ringrose2020store}. To ensure safe operations of
CO\textsubscript{2} storage, we develop a quantitative leakage detection
tool based on a deep neural classifier. This classifier is trained on
time-lapse images that contain the imprint of CO\textsubscript{2} plumes
that behave regularly and irregularly. In case of irregular flow,
CO\textsubscript{2} escapes the storage complex through a pressure
induced opening in the seal, which causes a localized increase in
permeability (shown in Figure~\ref{fig:diff_JRM}).

Because time-lapse differences are small in amplitude, and strongly
localized laterally when leakage occurs, highly sensitive learned
classifiers are needed. For this purpose, we follow
\citet{erdinc2022AAAIdcc} and adopt the Vision Transformer (ViT)
\citep{dosovitskiy2020image}. This state-of-the-art classifier
originated from the field of natural language processing (NLP)
\citep{vaswani2017attention}. Thanks to their attention mechanism, ViTs
have been shown to achieve superior performance on image classification
tasks where image patches are considered as word tokens by the
transformer network. As a result, ViTs have much less image-specific
inductive bias compared to convolutional neural networks
\citep{dosovitskiy2020image}.

To arrive at a practical and performant ViT classifier, we start from a
ViT that is pre-trained on image tasks with $16\times 16$ patches and
apply transfer learning \citep{yosinski2014transferable} to fine-tune
this network on $1576$ labeled time-lapse images. Catastrophic
forgetting is avoided by freezing the initial layers, which are
responsible for feature extraction, during the initial training. After
the initial training of the last dense layers, all network weights are
updated for several epochs while keeping the learning rate small. The
labeled (regular vs.~irregular flow) training set itself consists of
$1576$ time-lapse datasets divided equally between regular and irregular
flow.

\begin{figure}
\centering
\includegraphics[width=0.980\hsize]{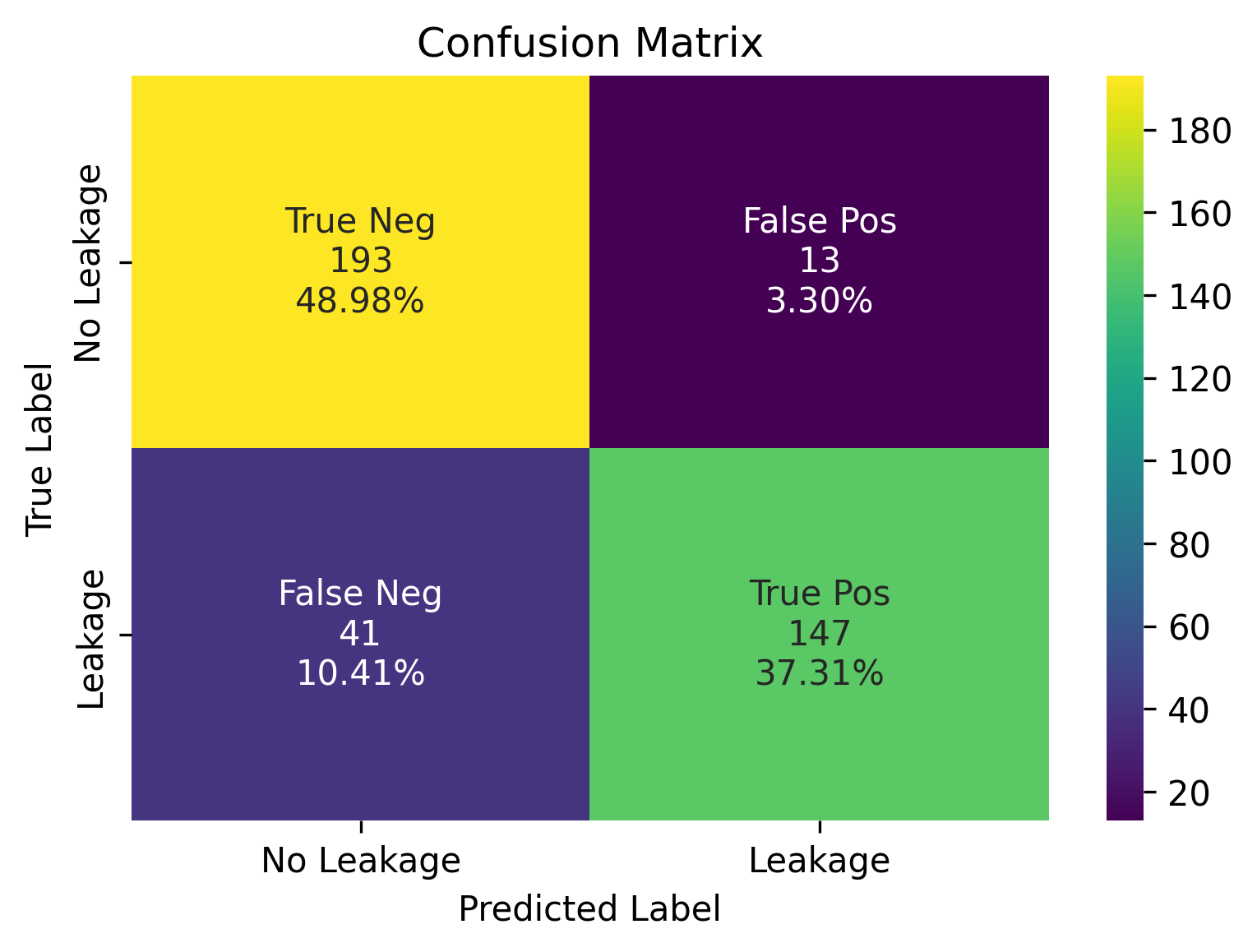}
\caption{Confusion matrix for classifier trained on recovery images from
JRM.}\label{fig-confusion}
\end{figure}

After the training is completed, baseline and monitor surveys are
simulated for $394$ unseen Earth models with regular and irregular
plumes. These simulated time-lapse datasets are imaged with JRM by
inverting the matrix in Equation~\ref{eq-jrm} via Bregman iterations in
Equation~\ref{LBk}. The resulting time-lapse difference images (see
Figures~\ref{fig:diff_JRM} and~\ref{fig:diff_JRM_noleak} for two
examples) serve as input to the ViT classifier. Refer to
Figure~\ref{fig-confusion} for performance, which corresponds to a two
by two confusion matrix. The first row denotes the classification
results for samples with regular plume (negative samples), where $193$
(true negative) out of $206$ samples are classified correctly. The
second row denotes the classification results for samples with
CO\textsubscript{2} leakage over the seal (positive samples), where
$147$ (true positive) out of $188$ samples are classified correctly. Due
to the fact that JRM recovers relatively artifact-free time-lapse
differences, the classifier does not pick up too many artifacts related
to finite acquisition as CO\textsubscript{2} leakage. This leads to much
fewer false alarms for CO\textsubscript{2} leakage.

\section{Class activation mapping based saliency
map}\label{class-activation-mapping-based-saliency-map}

\begin{figure}
\centering
\subfloat[\label{fig:cam-leak}]{\includegraphics[width=0.490\hsize]{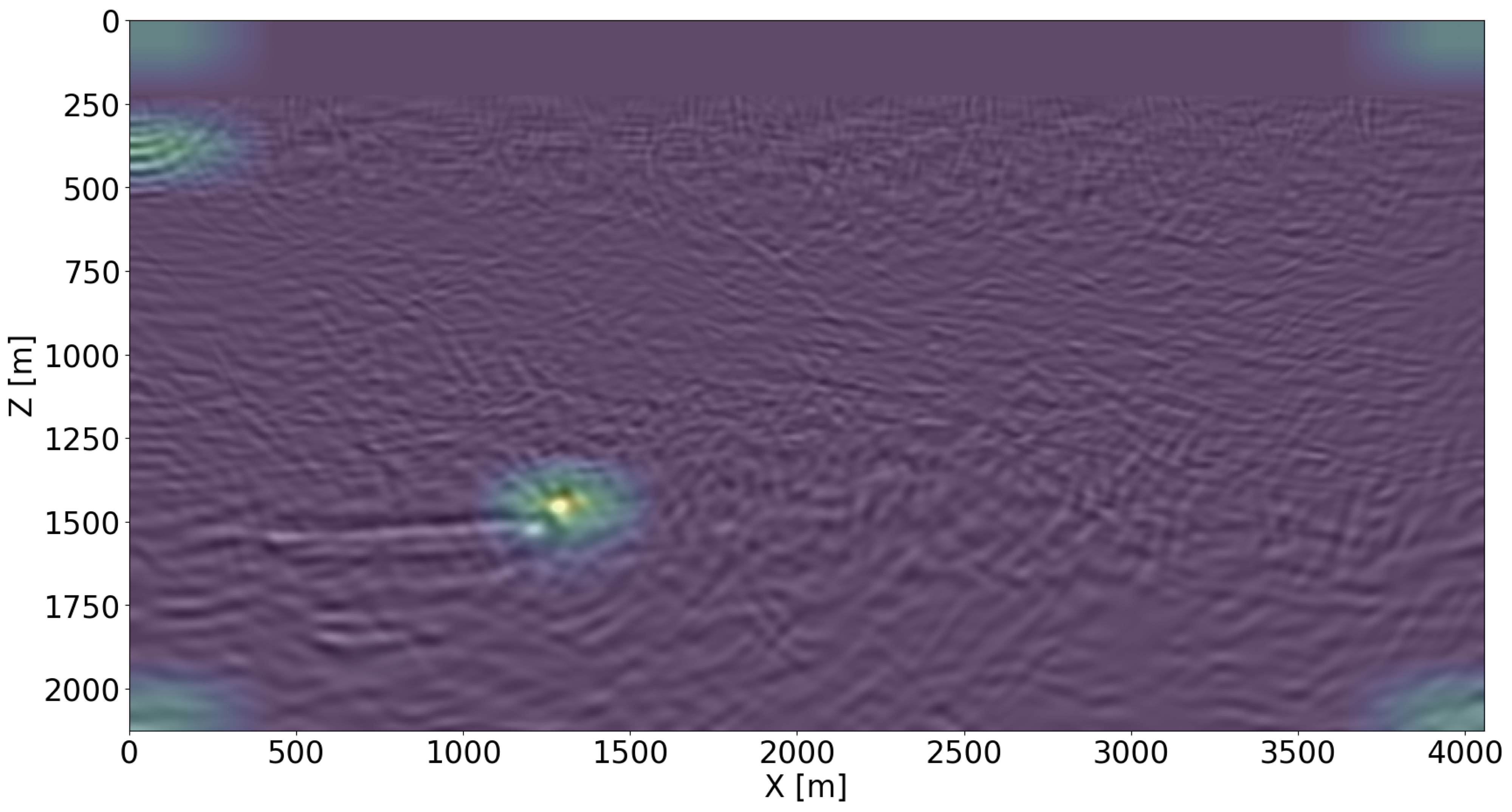}}
\subfloat[\label{fig:cam-noleak}]{\includegraphics[width=0.490\hsize]{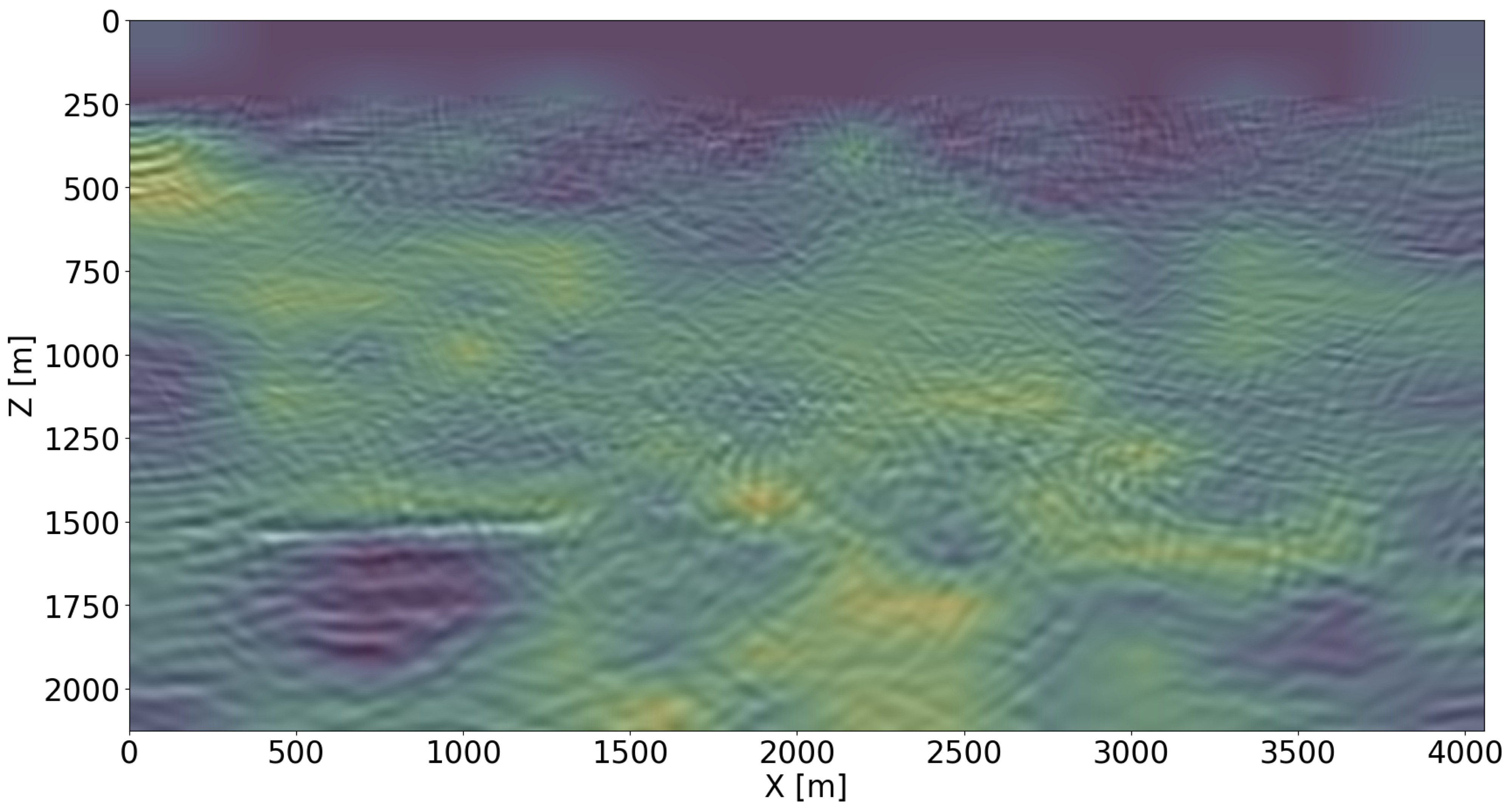}}
\caption{CAM for time-lapse difference images with a leaking plume and
with a regular plume.}\label{fig-cam}
\end{figure}

While our ViT classifier is capable of achieving good performance (see
Figure~\ref{fig-confusion}), making intervention decisions during GCS
projects calls for interpretability and trustworthiness of our
classifier
\citep{hooker2019benchmark, zhang2021survey, mackowiak2021generative}.
To enhance these features, we take advantage of class activation
mappings (CAM) \citep{zhou2016learning}. These saliency maps help us to
identify the discriminative spatial regions in each image that support a
particular class decision. In our application, these regions correspond
to areas where the classifier deems the CO\textsubscript{2} plume to
behave irregularly (if the classification result is leakage). By
overlaying time-lapse difference images with these maps, interpretation
is facilitated, assisting practitioners to make decisions on how to
proceed with GCS projects and take associated actions.
Figure~\ref{fig-cam} illustrates how the Score CAM approach
\citep{wang2020score} serves this purpose\footnote{We used the
  open-source software
  \href{https://github.com/jacobgil/pytorch-grad-cam}{PyTorch library
  for CAM methods} \citep{jacobgilpytorchcam} to calculate the CAM
  images.}. Figure~\ref{fig:cam-leak} shows the CAM result for a
time-lapse difference image classified as a CO\textsubscript{2} leakage
(in Figure~\ref{fig:diff_JRM}). Despite few artifacts around the image,
the CAM clearly focuses on the CO\textsubscript{2} leakage over the
seal, which could potentially alert the practitioners of GCS. When the
plume is detected as growing regularly, the CAM result is diffusive
(shown in Figure~\ref{fig:cam-noleak}). This shows that the
classification decision is based on the entire image and not only at the
plume area. The scripts to reproduce the experiments are available on
the SLIM GitHub page \url{https://github.com/slimgroup/GCS-CAM}.

\section{Conclusions \& discussion}\label{conclusions-discussion}

By means of carefully designed time-lapse seismic experiments, we have
shown that highly repeatable, high resolution and high fidelity images
are achievable without insisting on replication of the baseline and
monitor surveys. Because our method relies on a joint inversion
methodology, it also averts labor-intensive 4D processing. Aside from
establishing our claim of relaxing the need for replication empirically,
through hundreds of time-lapse experiments yielding significant
improvements in NMRS values, we also showed that a deep neural
classifier can be trained to detect CO\textsubscript{2} leakage
automatically. While the classification results are encouraging, there
are still false negatives. We argue that this may be acceptable since
decisions to stop injection of CO\textsubscript{2} are also based on
other sources of information such as pressure drop at the wellhead. In
future work, we plan to extend this methodology to different leakage
scenarios and quantification of uncertainty.

\section{Acknowledgement}\label{acknowledgement}

We would like to thank Charles Jones and Philipp A. Witte for the
constructive discussion. The CCS project information is taken from the
Strategic UK CCS Storage Appraisal Project, funded by DECC, commissioned
by the ETI and delivered by Pale Blue Dot Energy, Axis Well Technology
and Costain. The information contains copyright information licensed
under ETI Open Licence. This research was carried out with the support
of Georgia Research Alliance and partners of the ML4Seismic Center.

\bibliography{paper}

\end{document}